\documentclass[showpacs,twocolumn,superscriptaddress,nofootinbib]{revtex4-1} 

\usepackage[utf8]{inputenc}
\usepackage{hyperref}
\hypersetup{colorlinks=true,linkcolor=blue,citecolor=cyan}
\usepackage{graphicx}
\usepackage{xcolor, soul}
\sethlcolor{green}
\usepackage{dcolumn}
\usepackage{bm}
\usepackage{color}
\usepackage{enumitem}
\usepackage{mathrsfs}
\usepackage{amsmath}
\usepackage{amssymb}
\usepackage{cleveref}

\usepackage{soul,xcolor}
\setstcolor{red}

\begin{document}

\title{Energy extraction from a rotating Buchdahl star via magnetic reconnection}

\author{Ikhtiyor Eshtursunov}
\email{eshtursunovikhtiyor@gmail.com}
\affiliation{New Uzbekistan University, Movarounnahr str. 1, Tashkent 100000, Uzbekistan}

\author{Sanjar Shaymatov}
\email{sanjar@astrin.uz}
\affiliation{Institute of Fundamental and Applied Research, National Research University TIIAME, Kori Niyoziy 39, Tashkent 100000, Uzbekistan}
\affiliation{Tashkent University of Applied Sciences, Gavhar Str. 1, Tashkent 100149, Uzbekistan}
\affiliation{Tashkent State Technical University, 100095 Tashkent, Uzbekistan}

\date{\today}
\begin{abstract}
In this work, we investigate the magnetic reconnection (MR) process as a mechanism for energy extraction from a rapidly rotating Buchdahl star (BS), one of the most compact horizonless objects that can, in principle, possess a spin parameter exceeding the extremal limit of a black hole (BH). We explore the energetics of the BS by focusing on the newly proposed MR mechanism developed by Comisso and Asenjo (the Comisso–Asenjo mechanism). Within this framework, we evaluate the energy extraction efficiency and the associated power output from a rapidly rotating BS. We show that the ergoregion of the BS exists only when the spin parameter satisfies $\beta>1/\sqrt{2}$. Consequently, the extraction of rotational energy through MR becomes possible only under this condition. Furthermore, we analyze the rate of energy extraction driven by fast magnetic reconnection and compare the resulting power with that predicted by the Blandford–Znajek mechanism. Our results indicate that the energy extraction rate increases significantly when the BS spin parameter exceeds the extremal limit for a BH, highlighting that MR can be substantially more efficient than the Blandford–Znajek mechanism. We demonstrate that MR can greatly enhance energy extraction efficiency from rapidly rotating BS with a large spin, making such an object potentially more efficient engines of high-energy astrophysical processes than BH.

\end{abstract}
\pacs{}
\maketitle

\section{Introduction}
\label{introduction}

Recent observations, including gravitational-wave detections by LIGO–Virgo and EHT imaging of supermassive BHs \cite{Abbott16a,Abbott16b,Akiyama19L1,2022ApJ...930L..12E}, together with precise measurements of stellar orbits near the Galactic center \cite{2008ApJ...689.1044G,2010RvMP...82.3121G}, provide strong evidence for BHs. Despite these successes, the current observational accuracy still allows room for various alternatives to BHs, including horizonless compact objects. This highlights the need for robust theoretical and observational methods capable of distinguishing such alternatives from BHs, which play a fundamental role in explaining highly energetic astrophysical phenomena. Highly energetic astrophysical phenomena, including active galactic nuclei (AGN), gamma-ray bursts, and ultraluminous X-ray binaries \cite{Peterson:97book,Meszaros06,King01ApJ}, release enormous amounts of energy. Possible signatures of BH alternatives may arise from deviations of the spacetime geometry from the Kerr metric, although small deviations can be difficult to detect observationally. Horizonless compact objects may instead exhibit distinct magnetic field topologies near the horizon scale, since objects with a physical surface can sustain currents and generate intrinsic dipolar or multipolar fields, unlike BHs constrained by the no-hair theorem, which restricts the magnetic field to external configurations. Polarimetric observations together with non-thermal emission models already provide constraints on the magnetic field structure around BH candidates \cite{EventHorizonTelescope:2021bee,EventHorizonTelescope:2021srq}, offering a potential way to distinguish BHs from their alternatives. 

Recently, the Buchdahl star (BS) has been introduced as a compact object that saturates the Buchdahl compactness bound, $M/R \leq 4/9$, where $M$ and $R$ denote the mass and radius of the object (see details, e.g., \cite{Dadhich22,Alho22PRD,Shaymatov23PLB,Shaymatov23JCAP}). It is characterized by the gravitational potential $\Phi(r) = 4/9$, where $\Phi(r)$ represents the potential experienced by a radially infalling particle in a static/rotating spacetime. For comparison, BHs correspond to $\Phi(r) = 1/2$; thus, BSs represent extremely compact horizonless objects that closely approach the BH limit. It should be noted that the Kerr spacetime metric describes a rotating BH and does not, in general, represent a rotating non-BH compact object. Although static spacetime metrics such as the Schwarzschild or Reissner-Nordstr\"{o}m metric can represent both BH and non-BH objects, this distinction does not extend to stationary, axially symmetric spacetimes: the Kerr solution uniquely describes a rotating BH with spherical horizon topology once all higher multipole moments are fixed by the no-hair property. In contrast, a rotating BS possesses a non-null timelike boundary and may retain additional multipole structure. Since no exact solution exists for rotating non-BH compact objects, we adopt the Kerr spacetime metric as an approximation, justified by the fact that BSs are extremely compact configurations, characterized by $\Phi(R) = 4/9$, which lies close to the BH value $\Phi(R) = 1/2$. Under this approximation, we employ the Kerr metric to model a rotating BS. Interestingly, a BS can be sufficiently compact to form an ergosphere just above its surface for a large rotating parameter $a$. The extremal limit for a BS occurs at $a/M=9/8>1$, making it over-extremal relative to a BH. Consequently, a BS can maintain a greater rotation than the corresponding BH \cite{Dadhich22}. This provides additional motivation to investigate its energetics.

Energy extraction from rotating black holes (BHs) is one of the most intriguing predictions of general relativity. Penrose first showed that particle disintegration in the ergosphere can lead to energy extraction from a rotating BH \cite{Penrose:1969pc}. This mechanism and its extensions to the magnetic Penrose process (MPP) \cite{Bhat85,Parthasarathy86ApJ}, as well as the Blandford-Znajek (BZ) process \cite{Blandford1977} have been widely studied to explain the extreme luminosities of active galactic nuclei, microquasars, and gamma-ray bursts \cite{Rees:1984si,Peterson:97book,Meszaros06,King01ApJ,McKinney07}. Magnetic fields are crucial for energy extraction from rotating black holes. The Blandford–Znajek mechanism describes how magnetic fields threading the accretion disk can extract rotational energy from black holes \cite{Blandford1977,McKinney07}. This mechanism has since been investigated in numerous astrophysical settings \cite{Wagh89,Alic12ApJ,Moesta12ApJ}. In particular, the MPP extracts rotational energy via frame dragging and electromagnetic interactions, without the complex magnetohydrodynamic structure required by the Blandford–Znajek process. Therefore, the MPP \cite{Abdujabbarov11,Dhurandhar1984PRD,Dhurandhar1984PRD.30.1625, Wagh85ApJ,Tursunov:2019oiq,Shaymatov24PRD.110d4042S,Xamidov24EPJC} is particularly noteworthy for its exceptional efficiency. In various magnetic field configurations, the MPP can achieve efficiencies exceeding 100\% for charged particles \cite{Dadhich18MNRAS,Shaymatov24EPJC,Shaymatov22b,Khamidov25JCAP...03..053X}.

Magnetic reconnection (MR) near the horizon of rotating black holes has recently been proposed as a mechanism for high-energy particle acceleration. Frame dragging can twist magnetic field lines, producing antiparallel configurations in the equatorial plane and enabling MR within the ergosphere. In this process, some particles acquire negative energy and fall into the black hole, while others escape with enhanced positive energy, effectively extracting rotational energy. Early studies suggested that slow MR is inefficient \cite{Koide08ApJ}, whereas relativistic MR may generate energetic outflows comparable to the Blandford–Znajek mechanism \cite{Parfrey19PRL}. However, a detailed estimate of the energy released through MR remains to be established. More recently, Comisso and Asenjo \cite{Comisso21} proposed a different approach to energy extraction through MR in the vicinity of a rapidly rotating Kerr BH. Their work introduced a novel mechanism that allows the efficiency and power of energy extraction driven by MR to be quantitatively evaluated. They demonstrated that this process can operate as an efficient energy-extraction mechanism, with both efficiency and power strongly dependent on the BH spin parameter. This mechanism, now widely referred to as the Comisso–Asenjo mechanism, has since been applied in several studies of rapidly rotating BHs \cite{Liu22ApJ,Wei22,Carleo22,Khodadi22,Wang22}. The Comisso–Asenjo mechanism has also been applied to rotating MOG BHs, demonstrating the combined effect of the BH charge and the MOG parameter on magnetic-reconnection-driven energy extraction and interpreting the MOG parameter as an attractive gravitational charge \cite{Khodadi23MR_JCAP,Shaymatov24MR}. Further studies have explored energy extraction from BHs driven by MR, providing additional insight into diverse extraction mechanisms in different BH models \cite{Zhang24JCAP...07..042Z,Chen24PRD.110f3003C,Shen25PRD.111b3003S,Long25EPJC...85...26L,Rodriguez25PDU,Zeng25PRD.112f4032Z,Zeng25PRD.112f4080Z,Cheng25EPJC...85.1130C,Wang25JCAP,YuChih25PRD.112j4016Y}. In this process, the twisting and reconnection of magnetic field lines accelerate particles to relativistic energies, producing powerful jets from the BH poles \cite{Blandford:1982di}. These jets are believed to contribute to the production of ultra-high-energy cosmic rays and gamma-ray bursts \cite{Marscher:2008aa,Kotera2011ARA&A,Romero:2008zj}, making this mechanism a leading candidate for powering some of the most energetic phenomena in the Universe. 

In this work, we employ the Comisso–Asenjo mechanism to investigate energy extraction from a BS driven by MR. We analyze the efficiency and power of energy extraction as functions of the spin parameter, reconnection location, plasma magnetization, and magnetic field orientation. Our results reveal distinct features relative to the Kerr case and provide new insights into MR-driven processes, including ultra-high-energy cosmic ray acceleration. The structure of this paper is organized as follows: In Sec.~\ref{Sec:BS}, we present the conditions required for the existence of the ergosphere around the BS and briefly discuss the relevant properties of the Kerr spacetime used to model the rotating BS geometry and its associated physical characteristics. In Sec.~\ref{Sec:BS_MR}, we investigate the MR process in the vicinity of the BS. This is followed by an analysis of the power output, energy efficiency, and energy extraction rate associated with the MR mechanism, where we present our findings and provide a comparative discussion in Sec.~\ref{Sec:BS-Power-MR}. Finally, our concluding remarks are summarized in Sec.~\ref{Sec:con}. Throughout this work, we use a system of geometric units in which $G_{\rm{N}}=c=1$

\section{A rotating Buchdahl star metric }\label{Sec:BS}

Under the approximation mentioned, we model a rotating BS using the Kerr metric. This approximation is justified by compactness, characterized by the defining condition of the potential $\Phi(r)=4/9$ for BS, compared to $\Phi(r)=1/2$ for BH. Accordingly, we use the Kerr metric in the standard Boyer–Lindquist coordinates to describe the rotating BS, which reads as follows:
\begin{eqnarray}\label{eq:metric}
ds^2 &=&-\left(\frac{\Delta-a^2\sin^2\theta}{\Sigma}\right)dt^2 +\frac{\Sigma}{\Delta}dr^2+\Sigma d\theta^2\nonumber\\
&-&\frac{2a\sin^2\theta(r^2+a^2-\Delta)}{\Sigma}dtd\phi \nonumber\\
&+&\frac{(r^2+a^2)^2-\Delta a^2\sin^2\theta}{\Sigma}\sin^2\theta
d\phi^2 \, ,
\end{eqnarray}
with $\Sigma=r^2+a^2\cos^2\theta$ and $\Delta=r^2+a^2-2Mr$, where $a$ and $M$, respectively, refer to the spin and mass parameter of BS.  For this given spacetime, the gravitational potential reads as  
\begin{eqnarray}
\Phi(R) =\frac{M/r}{1+\beta^2 (M/r)^2}\, ,
\end{eqnarray} 
where we have used $\beta^2=a^2/M^2$. As mentioned, the defining condition of the potential is written as $\Phi(r)=4/9$ for BS and $\Phi(r) = 1/2$ for BH \cite{Dadhich20:JCAP}. It should be emphasized that the Buchdahl compactness bound is defined by $\Phi(r)\leq 4/9$ (see details, e.g., \cite{Dadhich22,Alho22PRD,Shaymatov23JCAP,Dadhich20:JCAP}), which allows one to define BS potential as follows:   
 \begin{eqnarray}
\Phi(r) = \frac{M/r}{1+\beta^2 (M/r)^2} = 4/9\, ,
\end{eqnarray}
from which one can define the BS surface radius 
\begin{eqnarray} \label{eq:rBS}
r_{BS}(\beta) =\frac{9M}{8}\left(1+\sqrt{1-(8/9)^2{\beta^2}}\right)\, .
\end{eqnarray}
For BH, it is defined by the horizon $r_{+}(\beta)= M(1+ \sqrt{1-\beta^2})$. 

\begin{figure}
    \centering
    \includegraphics[scale=0.6]{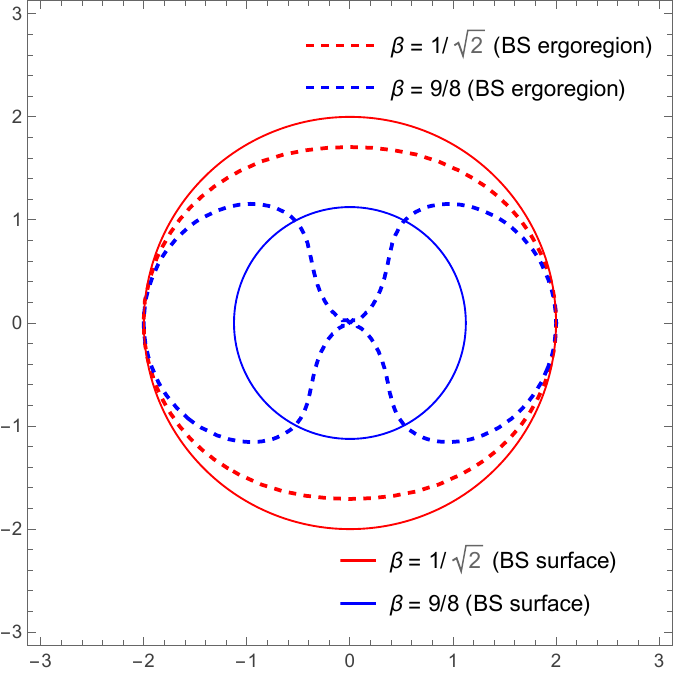}
\caption{\label{fig:ergo} The ergoregion (dashed) and the surface of BS (solid) for two different values of the spin parameter $\beta=1/\sqrt{2},\, 9/8$. }
\end{figure}

The Kerr spacetime possesses two Killing vectors, $\xi^{\alpha}_{(t)}=(\partial/\partial t)^{\alpha}$ and
$\xi^{\alpha}_{(\varphi)}=(\partial/\partial \phi)^{\alpha}$, associated with stationarity and axial symmetry, leading to conserved energy and angular momentum for test particles. The static surface is defined by the condition $g_{tt}= 0$, where the timelike Killing vector  $\xi^{\mu}_{(t)} = \partial/\partial t$ becomes null. 
\begin{eqnarray}
r_{st}(\beta,\theta)=M\left(1+\sqrt{1-\beta^2\cos^2\theta}\right)\, ,
\end{eqnarray}
which is referred to as the static surface, at which or below this any test particle cannot remain static. The ergosphere, bounded by the static surface $r_{st}$ and the horizon $r_{+}$, allows negative-energy particle states, enabling the MR process. For a BS, it extends between $r_{BS}$ and $r_{st}$, and is narrower than in the BH case since $r_{BS} > r_{+}$, as exhibited in Fig.~\ref{fig:ergo}. At extremality, the ergosphere width is $r_{st} - r_{+} = 2M-M = M$ for BH $(\beta=1)$ and $r_{st} - r_{BS}= 2M - 9/8\,M = 7/8\, M$ for BS $(\beta=9/8)$. Interestingly, the ergosphere of BS is narrower than that of BH since $r_{BS} > r_{+}$. However, the extremal limit for BS occurs at $\beta=9/8 >1$, thus making it over-extremal relative to BH (see details, e.g., \cite{Shaymatov24EPJC}).

As illustrated in Fig.~\ref{fig:ergo}, the region defined by $r_{st} -r_{BS}$ corresponds to the ergosphere of the BS. Notably, this ergoregion disappears in the limit $\beta \to 1/\sqrt{2}$. The BS ergoregion exists only for $\beta>1/\sqrt{2}$. This represents a distinctive feature of rotating BS configurations: in contrast to BHs, the extraction of rotational energy from a rapidly rotating BS through proposed mechanisms becomes possible only when the condition $\beta>1/\sqrt{2}$ is satisfied. This is the main distinction from BH: the rotational energy of a rotating BS with the spin parameter, i.e., $\beta\leq 1/\sqrt2$, is not extractable or applicable for powering high-energy jets and highly energetic astrophysical phenomena \cite{Peterson:97book,Meszaros06,King01ApJ}. 

Motivated by the above considerations, we next wish to examine the magnetic reconnection process in the vicinity of the BS.

\section{Magnetic reconnection and energy extraction from Buchdahl star}\label{Sec:BS_MR}

\begin{figure*} 
\begin{center} \begin{tabular}{c c} \includegraphics[scale=0.6]{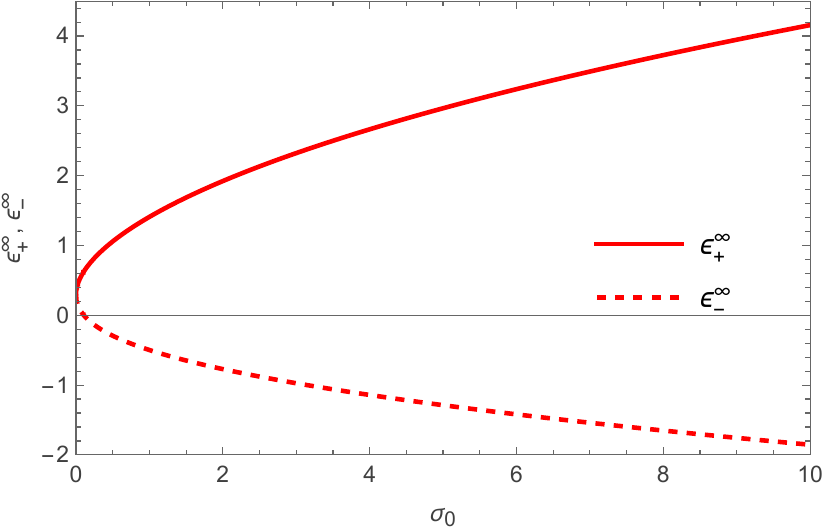}\hspace{1cm} \includegraphics[scale=0.6]{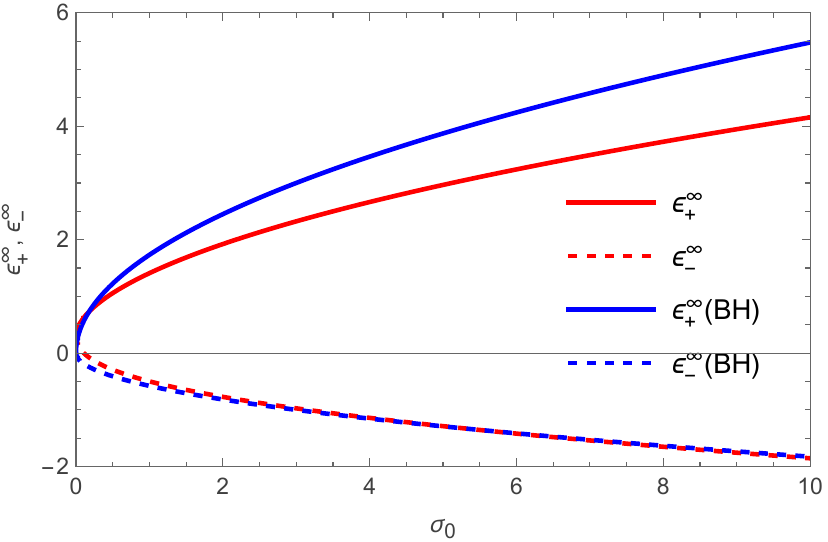}\hspace{0cm} 
\end{tabular} \caption{\label{Fig:energy} {The energies at infinity $\epsilon_{+}^{\infty}$ and $\epsilon_{-}^{\infty}$ per enthalpy are showed for BS and Kerr BH. The left panel: BS's $\epsilon_{+}^{\infty}$ (red line) and $\epsilon_{-}^{\infty}$ (red dashed line) are plotted for maximum energy extraction conditions: $\beta$, r/M$\to$9/8 and $\xi \to 0$. The right panel: $\epsilon_{+}^{\infty}$ (blue line) and $\epsilon_{-}^{\infty}$ (blue dashed line) are plotted for Kerr BH's maximal energy extraction conditions which are $\beta$, r/M$\to$1 and $\xi \to 0$ and compared with BS's case. }} 
\end{center}
\end{figure*}
Theoretically, we can consider Buchdahl stars as extraordinarily powerful sources of energy. Unlike other compact objects, in BS there is no event horizon, that is why energy harnessing processes can be easier and also much more effective. Comisso-Asenjo mechanism~\cite{Comisso21} which is considered outstanding application for energy extraction by using the MR technique from maximally spinning black holes strongly has a connection with frame dragging effects along with the lines of magnetic field around them. In our BS case, we can apply this mechanism efficiently to extract rotating star's energy in its ergoregion by using MR's aspects. To apply the mechanism, it is much more convenient to use zero angular momentum observer (ZAMO) frame, so we analyse plasma's energy density in it. We express the line element in ZAMO frame like this:
  \begin{eqnarray}
      ds^2=-d\hat{t}^2+\sum_{i=1}^3(d\hat{x^i})^2=\eta_{\mu \nu}\hat{dx^{\mu}}\hat{dx^{\nu}}\, , 
  \end{eqnarray}
in here we express $d\hat{t}$ and $d\hat{x^i}$ as below: 
\begin{eqnarray}
d\hat{t}=\alpha dt \mbox{~~and~~} d\hat{x^i}=\sqrt{g_{ii}}dx^i-\alpha \beta^{i}dt\, ,
\end{eqnarray}
where $\alpha$ is considered the lapse function and $\beta^i=(0,0,\beta^{\phi})$ indicates the shift vector, they are expressed as 
\begin{eqnarray}
\alpha=\sqrt{-g_{tt}+\frac{g_{\phi t}^2}{g_{\phi \phi}}}  \mbox{~~and~~}\beta^{\phi}=\frac{\sqrt{g_{\phi\phi}} \omega^{\phi}}{\alpha}\, .
\end{eqnarray}
In the formula $\omega^{\phi}=-{g_{\phi t }}/{g_{\phi \phi}}$ is named the angular velocity of the frame dragging. In the ZAMO frame covariant and contravariant components of the vector
b can be written as  
\begin{eqnarray}\label{Boyer}
\hat{b_0}=\frac{b_0}{\alpha}+\sum_{i=1}^3 \frac{\beta^i}{g_{ii}}b_i \mbox{~~and~~} \hat{b_i}=\frac{b_i}{\sqrt{g_{ii}}}\, ,\\
\hat{b^0}=\alpha b^0 \mbox{~~and~~} \hat{b^i}=\sqrt{g_{ii}}b^i-\alpha \beta^i b^0\, .
\end{eqnarray}
 We use the essential capability of MR mechanism to harness BS energy through analysing several conditions for the negative energy formation at infinity and going out to infinity of the accelerated and decelerated plasma with the use of reconnection process in the ergosphere region of the star. In our work, we do not have an interest to the origin of the plasma properties to provide accuracies, but we take a plasma with a given particle density and pressure properties. With using one-fluid approximation of the plasma, we can write energy-momentum tensor as 
\begin{eqnarray}
    T^{\mu \nu}=pg^{\mu \nu}+{\mathit{w}}U^{\mu}U^{\nu}+F^{\mu}_{\delta}F^{\nu \delta}-\frac{1}{4}g^{\mu \nu}F^{\rho \delta}F_{\rho \delta}\, .
\end{eqnarray}
In this equation, $p$ is the proper plasma pressure and $\mathit{w}$ refers to the density of enthalpy, while $U^{\mu}$ and $F^{\mu \nu}$ are four-velocity and the tensor of the electromagnetic field respectively. We should note that, here density of enthalpy is expressed as $\mathit{w}=e_{int}+p$, where the density of thermal energy is provided by \cite{Koide08ApJ} 
\begin{equation}
    e_{int}=\dfrac{p}{\Gamma-1}+\rho c^2\, ,
\end{equation}
where $\Gamma$ is adiabatic index and $\rho$ refers to the density of proper mass. We in the next expressions define relativistic hot plasma of BS with the equation of state by applying conditions we wrote above. We need energy density at infinity expression and it can be defined by the equation below
 \begin{eqnarray}
e^{\infty}=-\alpha g_{\mu 0}T^{\mu 0}\, .
\end{eqnarray}
We take this relation into account and write energy density at infinity like this:    
\begin{equation}
    e^{\infty}=\alpha \hat{e}+\alpha \beta^{\phi}\hat{P}^{\phi}\, ,
\end{equation}
in the equation $\hat{e}$ is considered total energy density and $\hat{P}^{\phi}$ defines the momentum density's azimuthal component, and their full expressions can be written as follows
\begin{eqnarray}
    \hat{e}={\mathit{w}}\hat{\gamma}^2-p+\frac{\hat{B}^2+\hat{E}^2}{2}\, ,\\
   \hat{P}^{\phi}={\mathit{w}}\hat{\gamma}^2\hat{v}^{\phi}+(\hat{B} \times \hat{E})^{\phi}\, ,
\end{eqnarray}
in here $\hat{v}^{\phi}$ is considered the plasma velocity's azimuthal component at the ZAMO frame. The Lorentz factor $\hat{\gamma}$ can be defined as 
\begin{equation}
\hat{\gamma}=\hat{U}^0=\sqrt{1-\sum_{i=1}^3(d\hat{v}^i)^2}\, ,\\ 
\end{equation}
furthermore, electric and magnetic field components $\hat{E}^i$ and $\hat{B}^i$ which appear equation above are defined as
\begin{eqnarray}
\hat{B}^i=\epsilon^{ijk}\hat{F}_{jk}/2 \mbox{~~and~~} \hat{E}^i=\eta^{ij}\hat{F}_{j0}=\hat{F}_{i0}\, .
\end{eqnarray}
We should note that, energy density at infinity $e^{\infty}$ includes two components: hydrodynamic and electromagnetic parts with $e^{\infty}=e_{hyd}^{\infty}+e_{em}^{\infty}$ and each of them can be written as separately as follows:
\begin{eqnarray}
 e_{hyd}^{\infty}=\alpha \hat{e}_{hyd}+\alpha \beta^{\phi}{\mathit{w}}\hat{\gamma}^2\hat{v}^{\phi}\, ,   \\
 e_{em}^{\infty}=\alpha \hat{e}_{em}+\alpha \beta^{\phi}(\hat{B}\times \hat{E})_{\phi}\, , 
\end{eqnarray}
in the equations above $\hat{e}_{hyd}={\mathit{w}}\hat{\gamma}^2-p$ and $\hat{e}_{em}=(\hat{B}^2+\hat{E}^2)/2$ are considered energy densities of the hydrodynamic and electromagnetic fields at the ZAMO frame.
\begin{figure*}
    \centering
    \includegraphics[scale=0.6]{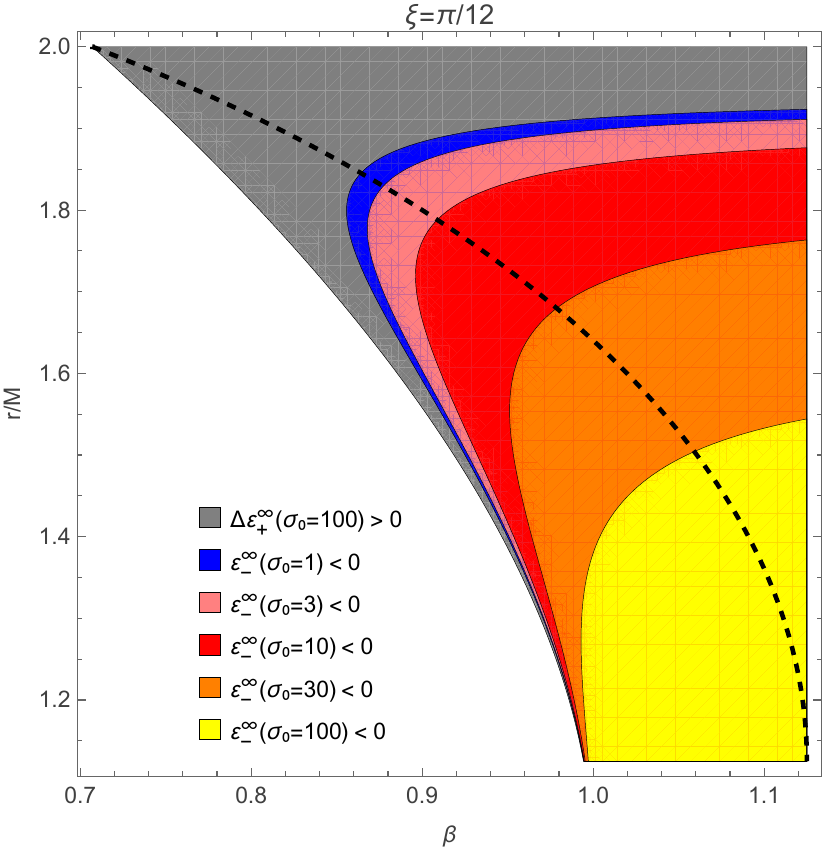}
    \includegraphics[scale=0.6]{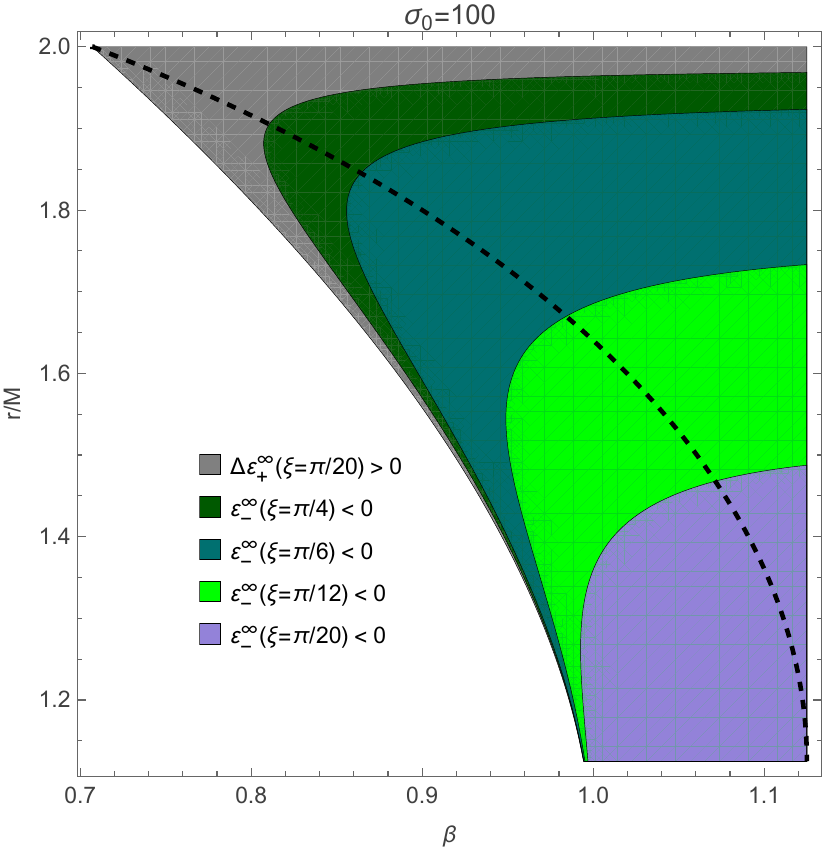}
    \caption{\label{fig:phase-space}Regions of the phase-space \{$\beta$, $r/M$\} are shown for the energy at infinity per enthalpy of accelerated plasma $\Delta\epsilon^{\infty}_{+}>0$ and the energy at infinity per enthalpy of decelerated plasma  $\epsilon^{\infty}_{-}<0$.  Left panel: \{$\beta$, $r/M$\} connection can be seen for $\Delta\epsilon^{\infty}_{+}>0$ (gray area) and $\epsilon^{\infty}_{-}<0$ (blue to yellow areas), for orientation angle $\xi=\pi/12$ and different values of the plasma magnetization parameter $\sigma_0$ $\in$ $\{$1,3,10,30,100$\}$ in the case of BS. Right panel: \{$\beta$, $r/M$\} connection is shown for $\Delta\epsilon^{\infty}_{+}>0$ (gray area) and $\epsilon^{\infty}_{-}<0$ (dark green to violet areas), for plasma magnetization parameter $\sigma_0=100$ and different values of the orientation angle $\xi$ $\in$ $\{$$\pi/20,\pi/12,\pi/6,\pi/4$$\}$ for BS case. Dashed black lines in both graphs represent BS surface radius.}
\end{figure*}
To define energy extraction via MR mechanism from BS, we should evaluate energy density at infinity. To do so, in comparison with the hydrodynamic energy density at infinity, the electromagnetic field's contribution can be eliminated due to its little impact on energy density at infinity. Nonetheless, it is crucial to remember that the bulk of magnetic field energy can be transformed into plasma kinetic energy in the MR mechanism. There should be taken all conditions into account and in our BS case, we take non-compressible and adiabatic plasma to do more approximation. That is why energy density at infinity expression can be written as following form~\cite{Comisso21}
   \begin{equation}
    e^{\infty}=e^{\infty}_{hyd}=\alpha [{\mathit{w}}\hat{\gamma}(1+\beta^{\phi}\hat{v}^{\phi})-\frac{ p}{\hat{\gamma}}]\ .
\end{equation}
Then, to analyse the process in a small scale and to make more localized, we should introduce the local rest frame. Our local rest frame is defined as $x'^{\mu}=(x'^{0},x'^{1},x'^{2},x'^{3})$ because we take this for bulk plasma which orbits with Keplerian angular velocity $\Omega_K$ at the equatorial plane around BS. We can write Keplerian angular velocity as follows
\begin{equation}\label{Eq:angular1}
    \Omega_K=\pm\frac{M^{1/2}}{r^{3/2}\pm aM^{3/2}}, .
\end{equation}
In our case, to choose the frame $x'^{\mu}$ we consider that the direction of $x'^{1}$ should be parallel to the radial direction $x^1=r$ and the direction of $x'^{3}$ should be parallel to the azimuthal direction $x^3=\phi$. At the ZAMO frame, we can write co-rotating Keplerian velocity by using Eq.~(\ref{Boyer}) as follows
\begin{eqnarray}
 \hat{v}_K&=&\frac{d\hat{x}^{\phi}}{d\hat{x}^{t}}
 =\frac{\sqrt{g_{\phi \phi}}dx^{\phi}/d\lambda-\alpha\beta^{\phi}dx^t/d\lambda}{\alpha dx^t/d\lambda} \nonumber \\
 &=&\frac{\sqrt{g_{\phi \phi}}}{\alpha}\Omega_K-\beta^{\phi}\, . 
 \end{eqnarray}
in this equation, we can obtain the forms of $\hat{v}_K$ and the Lorentz factor $\hat{\gamma}_K=1/\sqrt{1-\hat{v}_K^2}$ by using the Keplerian velocity formula $\Omega_K$ which was provided by Eq.~(\ref{Eq:angular1}). Applying MR mechanism for harnessing BS energy is highly connected with electromagnetic field properties and dynamics of plasma matter in the star. As we mentioned above, we consider adiabatic and non-compressible plasma approximations for one-fluid plasma, that is why hydrodynamic energy at infinity per enthalpy expression can be written as \cite{Comisso21}:
\begin{eqnarray}\label{Eq:essential}
 \epsilon^{\infty}_{\pm}&=&\alpha \hat{\gamma}_K \Bigg[(1+\beta^{\phi}\hat{v}_K)\sqrt{1+\sigma_0}\pm \cos{\xi}(\beta^{\phi}+\hat{v}_K)\sqrt{\sigma_0}\nonumber\\
 &-&\frac{\sqrt{1+\sigma_0}\mp \cos{\xi}\hat{v}_K\sqrt{\sigma_0}}{4\hat{\gamma}^2(1+\sigma_0-\cos^2{\xi}\hat{v}_K^2\sigma_0)}\Bigg]\, , 
\end{eqnarray}
here $\sigma_0=B_0^2/{\mathit{w}}$ is considered magnetization parameter of plasma and $\xi$ refers to the orientation angle between the outflow plasma directions at the equatorial plane and magnetic field. To analyse MR process for harnessing energy from BS, in plasma's accelerated state energy at infinity per enthalpy should be positive. If plasma is in a decelerated state near the BS surface, the energy at infinity per enthalpy will be negative, like explored in Penrose mechanism. That is why, our energy at infinity per enthalpy value should be positive to extract energy from BS and also should be much more greater than thermal energies. Furthermore, in the process, rest mass of plasma should be less than extracted one as energy to provide positiveness of the energy at infinity value. We consider the plasma which we are analysing is relativistically hot and it satisfies the equation of state condition, i.e., $\mathit{w} = 4p$ \cite{Comisso21}. Taking this into account, energies at infinity per enthalpy of the accelerated and decelerated plasma can be given as follows:
\begin{eqnarray}
 \epsilon_{-}^{\infty}<0 \mbox{~~and~~} \Delta \epsilon_{+}^{\infty}>0\, ,
 \end{eqnarray}
here $\Delta \epsilon_{+}^{\infty}$ can be expressed like 
 \begin{eqnarray}
 \Delta \epsilon_{+}^{\infty}=\epsilon_{+}-\left(1-\frac{\Gamma}{\Gamma-1}\frac{p}{\mathit{w}}\right)>0\, .
\end{eqnarray}
If we take $\Gamma$ polytropic index as 4/3, for relativistically hot plasma we can take this form  $ \Delta \epsilon_{+}^{\infty}=\epsilon_{+}^{\infty}>0$.

For discovering outstanding properties of energy extraction of BS with the use of magnetic reconnection mechanism we try to understand aspects of the accelerated and decelerated energies at infinity per enthalpy of plasma: $\epsilon_{+}^{\infty}$ and $\epsilon_{-}^{\infty}$. But, these expressions become very long and complicated if we try to define them analytically. That is why in Fig.~\ref{Fig:energy} we build the connection between energies $\epsilon_{+}^{\infty}$ and $\epsilon_{-}^{\infty}$ at infinity per enthalpy and magnetization parameter for maximum energy extraction conditions. In Fig.~\ref{Fig:energy}'s left panel, with BS's maximum energy extraction conditions by MR mechanism which are $\beta$,r/M $\to$9/8 and $\xi$$\to$0, energies at infinity $\epsilon_{+}^{\infty}$ and $\epsilon_{-}^{\infty}$ have relations with magnetization parameter $\sigma_0$ like cyclic loading behavior. This can define that the energies at infinity strongly depend on magnetic field values. In the right panel of Fig.~\ref{Fig:energy}, connection between energies at infinity per enthalpy $\epsilon_{+}^{\infty}$ and $\epsilon_{-}^{\infty}$ for BS and Kerr-rotating BH($\beta$,r/M $\to$1 and $\xi$$\to$0) in their maximum energy extraction regimes is plotted to compare with each other. It can be seen from the graph that energy at infinity for decelerated plasma $\epsilon_{-}^{\infty}$ for both compact objects faces nearly the same changes with the relation of magnetization parameter $\sigma_0$. However, energy at infinity for accelerated plasma $\epsilon_{+}^{\infty}$ changes more efficiently in the Kerr BH case than BS's.

 Furthermore, to know more about energy extraction properties of BS via MR process, in Fig.~\ref{fig:phase-space} we analyse phase-space regions $\{\beta, r/M\}$ where $\epsilon_{-}^{\infty}<0$ and $\Delta\epsilon_{+}^{\infty}>0$. In the left panel of Fig.~\ref{fig:phase-space} the relation is built for fixed magnetic field's orientation angle $\xi$=$\pi/12$ and magnetization parameter's different values $\sigma_0$ $\in$ $\{$1,3,10,30,100$\}$. In the right panel, the $\{\beta, r/M\}$ relation is made  for $\sigma_0$=100 and several different types of orientation angle $\xi$ $\in$ $\{$$\pi/20,\pi/12,\pi/6,\pi/4$$\}$. As can be seen from the left side graph of Fig.~\ref{fig:phase-space}, when plasma's magnetization becomes higher, $\{\beta, r/M\}$ phase-space region where MR harness BS energy extends to higher values of r/M and lower values of spin parameter $\beta$. We can see that $\epsilon_{-}^{\infty}$ decreases monotonically with the change of $\sigma_0$ and $\epsilon_{+}^{\infty}$ monotonically increases as a function of $\sigma_0$. For BS, r $<$$r_{BS}$,  $\frac{1}{\sqrt{2}}$ $<$ $\beta$ $<$ 9/8 and $\xi$ $<$ $\pi/2$ conditions satisfy $\epsilon^{\infty}_{+}>0$ and energy extraction can be effectively done. However, for $\epsilon^{\infty}_{-}<0$ condition, $\sigma_0$ $>>$ 1 should be done for energy extraction by reconnection from BS.

For energy extraction through MR technique, the orientation angle between the plasma flow directions at the BS's equatorial plane and magnetic field plays an essential role. $\{\beta, r/M\}$ phase-space area where the process of extracting energy takes place becomes higher to larger r/M values and lower $\beta$ values as the orientation angle $\xi$ becomes less. A decrease in the orientation angle $\xi$ leads to a higher energy extraction region because rotational energy extraction depends completely on the azimuthal component of the outflow velocity.
 
In the next section, we investigate MR process' power and energy extraction for BS.

\begin{figure*}
    \centering
    \includegraphics[scale=0.6]{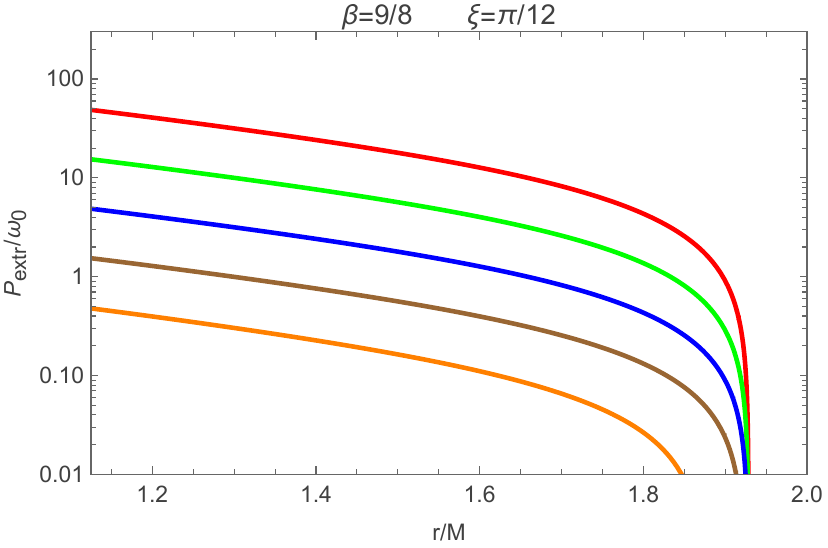}
    \includegraphics[scale=0.6]{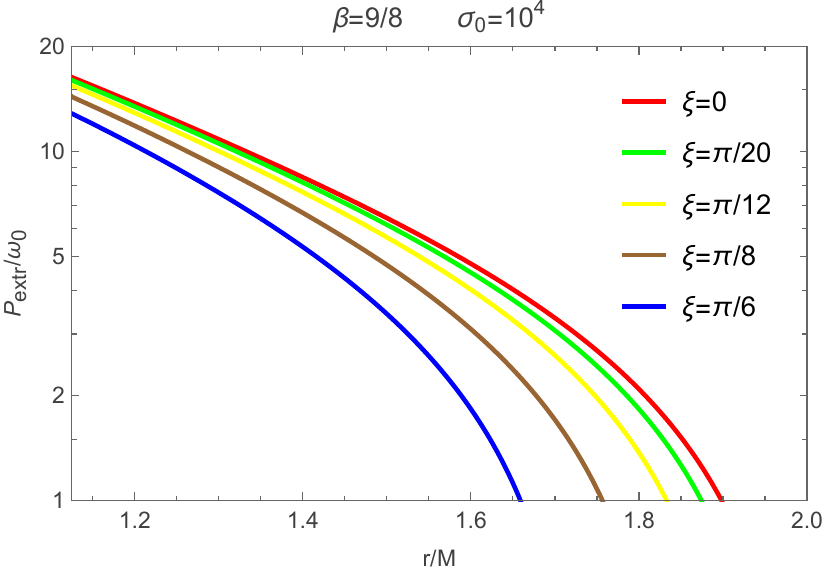}
    \caption{\label{fig:efficiency} Graphs show the connection between $P_{extr}=-\epsilon_{-}^{\infty}{\mathit{w}}_0 A_{in} U_{in}$ and r/M which is dominant X-point location for a rapidly rotating BS with spin parameter $\beta$=9/8. In the relation, MR inflow 4-velocity is $U_{in}=0.1$ which is in the regime of collisionless reconnection, and we showed $\epsilon_{-}^{\infty}$ above in Eq.(40) and in our BS case $A_{in}=(r_{BS}^2-r_{st}^2)$. We set M=1. Different colors from red to orange define different values of plasma magnetizations: from $\sigma_0$=10 to $\sigma_0$=$10^5$ and $\xi$=$\pi/12$ in the top panel. In the bottom panel, colors from red to blue refer to different orientation angles from $\xi$=$\pi/6$ to $\xi$=0 and $\sigma_0$=$10^4$.}
\end{figure*}

\section{\label{Sec:BS-Power-MR}THE POWER AND EFFICIENCY OF MAGNETIC RECONNECTION MECHANISM}

This section examines energy extraction from BS using the MR mechanism and assesses energy efficiency using the Comisso-Asenjo mechanism \cite{Comisso21}. We should note that, extracted energy and power strongly depends on BS's swallowed plasma with negative energy at infinity in the unit time. That is why how large the reconnection rate becomes, this makes energy extraction rate more. We initially write the expression of power of the escaping plasma $P_{extr}$ which separated from BS can be defined by \cite{Comisso21}
\begin{eqnarray}
    P_{extr}=-\epsilon_{-}^{\infty}{\mathit{w}}_0 A_{in} U_{in}\, ,
\end{eqnarray}
in the equation, $\epsilon_{-}^{\infty}$ is energy at infinity per enthalpy for decelerated plasma, $U_{in}$ depends on MR regime. For collisionless regime of the process it is taken as $U_{in}={\cal O}(10^{-1})$ and for collisional one $U_{in}={\cal O}(10^{-2})$. In our case, we take MR in collisionless regime and that is why $U_{in}={\cal O}(10^{-1})$ is satisfied. Furthermore, $A_{in}$ indicates the inflowing plasma's cross-sectional area and can be written as $A_{in}$$\sim$$(r_{st}^2-r_{BS}^2)$ for maximally rotating BS. In our case, $\beta$$\to$9/8 and $(r_{st}^2-r_{BS}^2)$ $\sim$ 2.73$M^2$. 

 We show in Fig.~\ref{fig:efficiency} the connection between r/M which is dominant X-point coordinates and $P_{extr}/{\mathit{w}}_0$ power extracted from maximally rotating BS with spin parameter $\beta$=9/8 in the MR mechanism's non-collisional regime. In the graph at the top panel, there plot is made for orientation angle of magnetic field while reconnecting $\xi$=$\pi/12$ and several different values of magnetization parameter $\sigma_0$. In the bottom panel, the relation is made for a fixed magnetization value $\sigma_0$=$10^4$ and different values of the angle of magnetic field orientation $\xi$. It can be noticed from the pictures that BS's extracted power increases monotonically for higher amounts of plasma magnetization and less values of the magnetic field orientation angle. The peak extractable power rises to a maximum as r/M and beta tend toward 9/8.
\begin{figure}
\centering
    \includegraphics[scale=0.6]{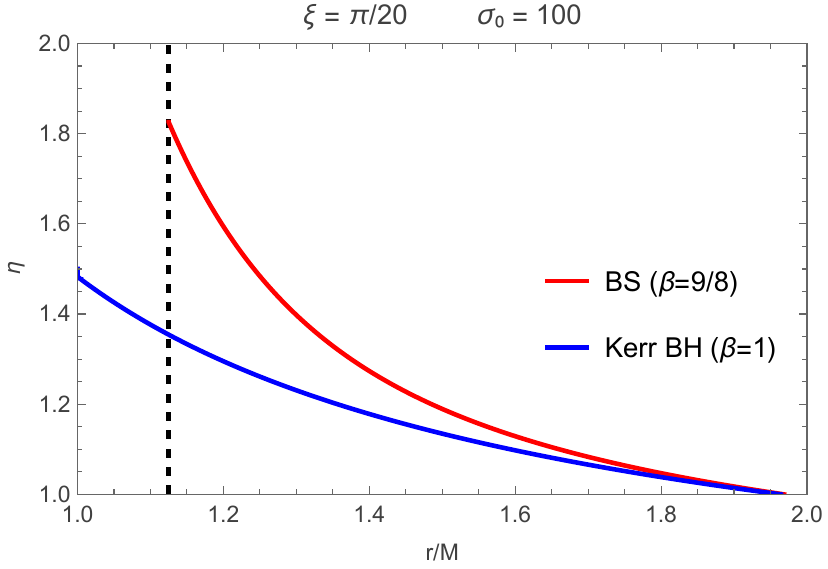}
    \caption{\label{fig:efficiencyeta} Connection between efficiency of MR process $\eta$ and as its function r/M which is dominant X-point coordinates for BS and Kerr BH in their spin parameters' conditions of maximal energy extraction, $\beta$=9/8 and $\beta$=1, respectively. The relation is built for magnetization's parameter of plasma $\sigma_0$=100 and orientation angle of magnetic field which is reconnection process occuring $\xi$=$\pi$/20. From dashed vertical line, BS surface radius $r_{BS}$ starts to exist}
\end{figure}

  Now we analyse another property of MR rate: energy extraction efficiency for BS. We should pay attention to energy harnessing process by this model to know how efficient is the mechanism. That is why it is interesting for us that how much energy can be harnessed by this process from BS. While the process is happening, magnetic energy is needed to create MR and generating plasmas with negative energy at infinity and plasmas escaping to infinity. In the mechanism, plasmas with negative energy at infinity fall to the BS and the ones with positive energies at infinity serve to extract star's energy. Therefore, plasma's energy extraction efficiency by this mechanism can be written as follows
\begin{eqnarray}
    \eta=\frac{\epsilon_{+}^{\infty}}{\epsilon_{+}^{\infty}+\epsilon_{-}^{\infty}}\, ,
\end{eqnarray}
in the relation $\epsilon_{+}^{\infty}$ and $\epsilon_{-}^{\infty}$ correspondingly refer to the accelerated and decelerated plasma 
energies at infinity per enthalpy, as mentioned above. It is essential to note that plasma energy can be extracted from BS when $\eta=\epsilon_{+}^{\infty}/(\epsilon_{+}^{\infty}+\epsilon_{-}^{\infty})>1$ condition satisfies for the efficiency of energy extraction. 

\begin{figure*}
    \centering
    \includegraphics[scale=0.45]{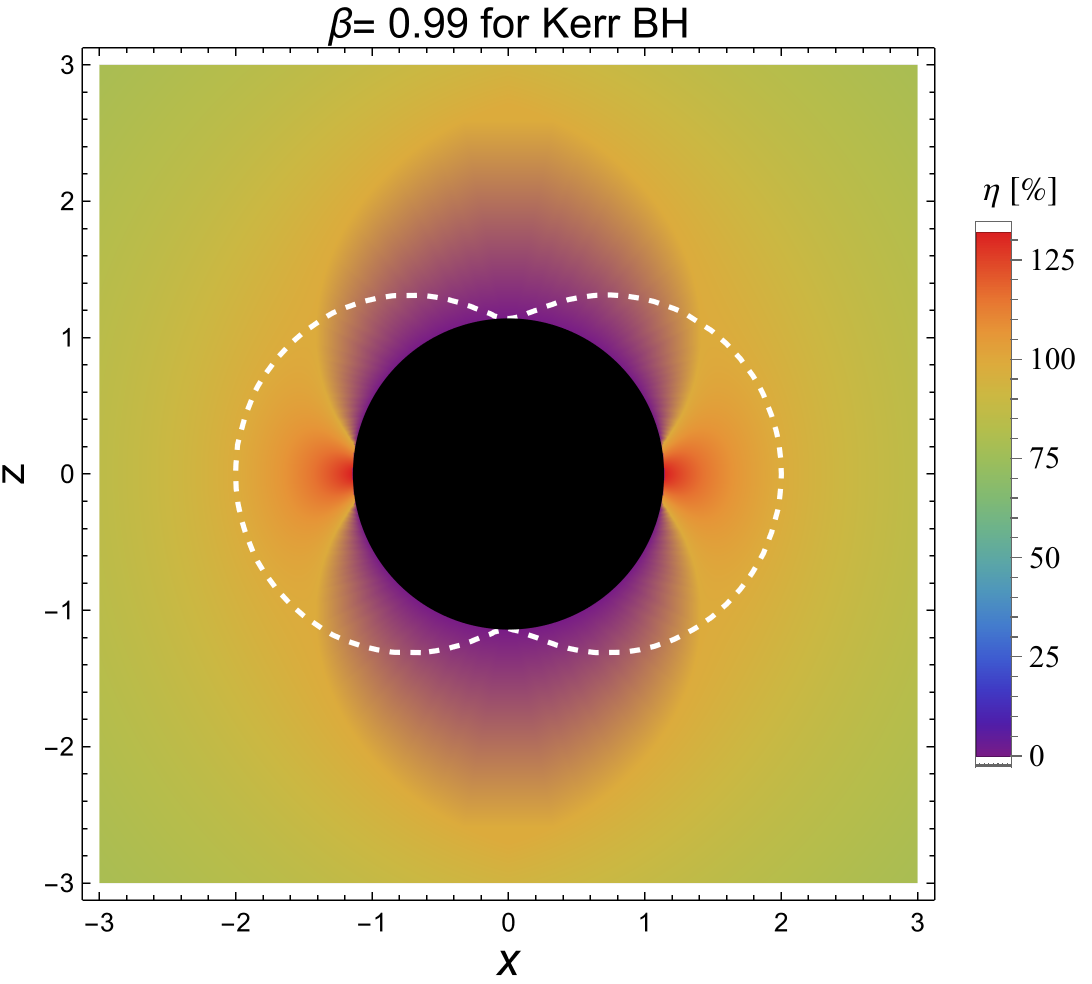}
    \includegraphics[scale=0.45]{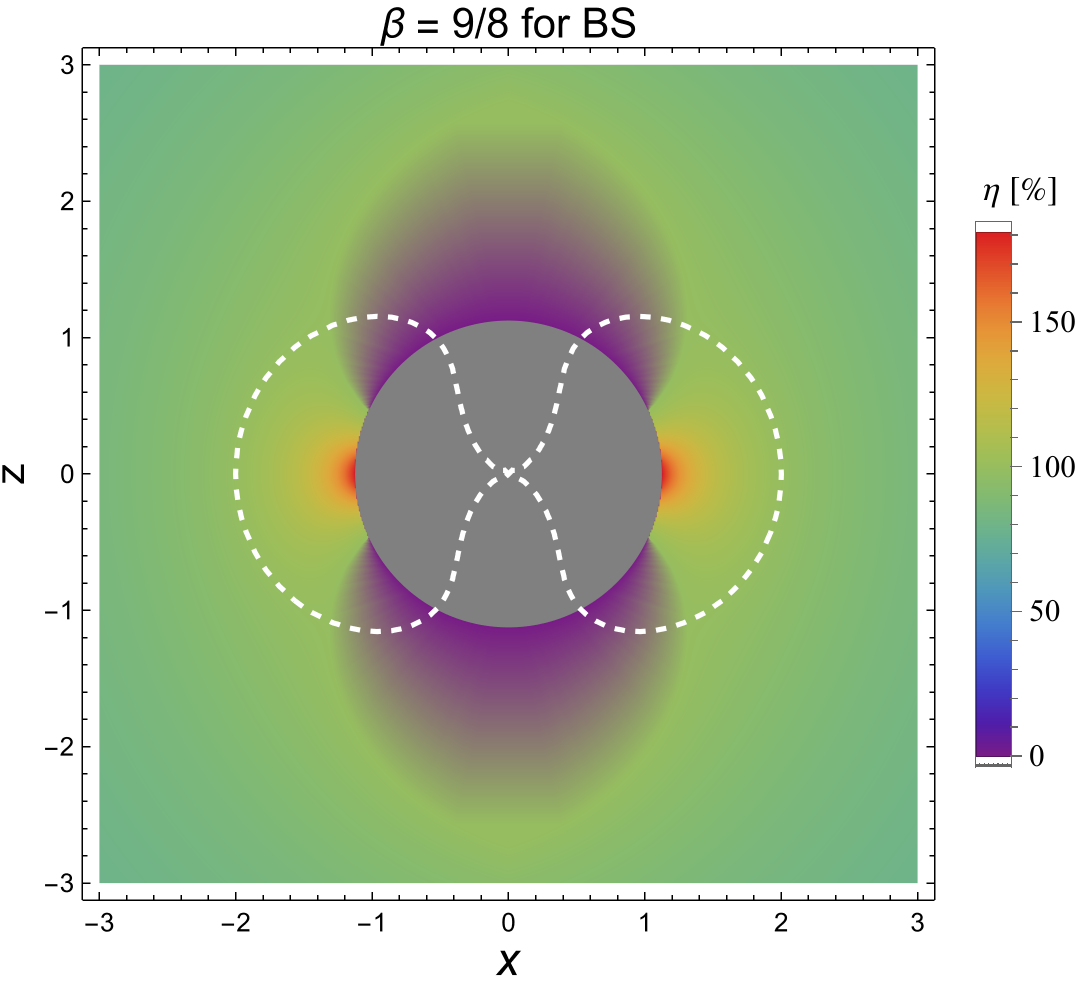}
    \caption{\label{fig:efficiencyplot} Energy efficiencies' plots of MR mechanism are given for Kerr BH and BS with spin parameters $\beta$=0.99 and $\beta$=9/8, respectively. Furthermore, these compact objects' ergospheres are showed around them. These graphs are provided for MR process' parameters $\sigma_0$=100 and $\xi$=$\pi/20$. }
\end{figure*}
\begin{figure}
    \centering
    \includegraphics[scale=0.6]{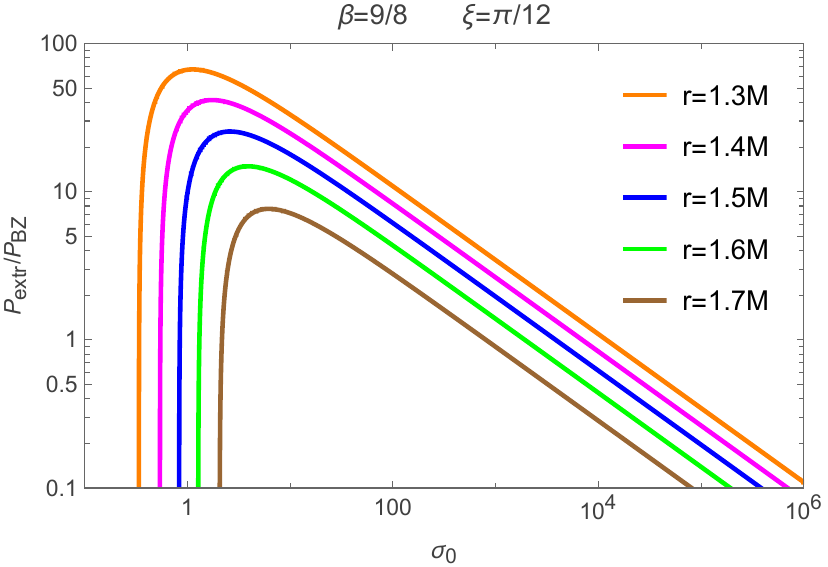}
    \caption{\label{fig:efficiencyratio} The relation between plasma magnetization parameter $\sigma_0$ and as its function $P_{extr}/P_{BZ}$ power ratio for BS with spin parameter $\beta$=9/8 and magnetic field's orientation angle which reconnection occurs $\xi$=$\pi/12$. In this figure, every different color(orange to brown) means different values of r/M which are X-dominant point coordinates. We take the regime as non-collisional that is why $U_{in}=0.1$, $A_{in}=(r_{st}^2-r_{BS}^2)$ and $\kappa$ refers to magnetic field geometry near BS and for our case $\kappa$=0.05 is used. }
\end{figure}
 In Fig.~\ref{fig:efficiencyeta} the relation is made for assessing and comparing MR method's energy extraction efficiencies for two different cases: Kerr BH and BS cases. The efficiency $\eta$ is given as a function of r/M which is considered dominant X-point coordinates with magnetic field orientation angle $\xi$=$\pi$/20 and magnetization parameter $\sigma_0$=100. For Kerr BH and BS, spin parameters are taken for maximal energy extraction condition's values, $\beta$=1 and $\beta$=9/8. Vertical dashed line refers to the starting points of BS surface radius. In both cases, the efficiencies rises noticeably for MR's X-coordinates are nearer to the BH event horizon and BS surface respectively. As we see in figure, the efficiency for BS becomes higher as result of greater spin and rotation effects.

 Now we make some comparison between two different methods of energy extraction from BS which uses magnetic field: MR mechanism and BZ process\cite{Blandford1977}. In BZ mechanism BS's rotational energy can be extracted electromagnetically by using magnetic field aspects. In BS case, energy extraction process' efficiency rate via the BZ mechanism can be written like Kerr BH's which was analysed in~\cite{tchekhovskoy_black_2010} 
 \begin{eqnarray}
   P_{BZ} \simeq \kappa \Phi_{BS}^2 (\Omega_{BS}^2+\chi\Omega_{BS}^4+\zeta\Omega_{BS}^6)
\end{eqnarray}
 in the expression, $\Phi_{BS}$ is magnetic flux spreading to BS surface, $\Omega_{BS}$=a/2$r_{BS}$ BS surface's angular frequency and $\kappa$, $\chi$ and $\zeta$ are considered constant values. $\kappa$ coefficient is choosen by analysing compact objects' geometry~\cite{pei_blandfordznajek_2016}, for our BS case, we choose $\kappa$$\approx$0.053 as a result of geometry near the star. And other constant values are $\chi$$\approx$1.38 and $\zeta$$\approx$-9.2. $P_{BZ}$$\simeq$$\kappa$ $\Phi_{BS}^2$$(a/4M)^2$ can be described by using BZ expanding for little values of spin a$\ll$ 1. 
 By using some analytical approximations in ~\cite{Comisso21}, we can compare the power extracted from BS by the plasma escape and power available via BZ process and write the ratio $P_{extr}/P_{BZ}$ as
 \begin{eqnarray}
\frac{P_{extr}}{P_{BZ}}\sim \frac{-\epsilon_{-}^{\infty} A_{in} U_{in}}{\kappa \Omega_{BS}^2r_{BS}^4\sigma_0\sin^2{\xi}(1+\chi\Omega_{BS}^2+\zeta\Omega_{BS}^4)}
\end{eqnarray}
In Fig.~\ref{fig:efficiencyplot} we described energy extraction efficiencies' distributions of MR mechanism for two different types of compact objects: Kerr BH and BS for their spin parameters' values $\beta$=0.99 and $\beta$=9/8, correspondingly. Around them, we can see their ergoregions. It is evident that the energy extraction efficiency of these compact objects through MR are roughly higher than 100 $\%$. Noticeably BS's efficiency is much greater because of the differences between two objects' rotation and surface radius boundaries. 

We can see from Fig.~\ref{fig:efficiencyratio} that, for the non-collisional MR process in BS, the connection is built for $P_{extr}/P_{BZ}$ division and as its function of magnetization value of the plasma $\sigma_0$.The height of the ratio of powers decreases and the curves move to the right to larger $\sigma_0$ as the location r/M increases. However, it does meet $P_{extr}/P_{BZ}>1$, it makes the BZ mechanism significantly less effective than the MR. However, as $P_{extr}/P_{BZ}\to0$ in this limit, energy extraction through fast MR is always subdominant to the BZ process if $\sigma_0$ grows to infinity.

\section{Conclusion}
\label{Sec:con}

Black holes are powerful energy reservoirs, and their extractable rotational energy is central to high-energy astrophysical phenomena. We consider a rapidly rotating BS, which is the possible most compact object without event horizon as an alternative candidate to black holes. We showed that ergosphere formation occurs only for $\beta >1/\sqrt{2}$, establishing a threshold for energy extraction. For $\beta \leq 1/\sqrt{2}$, the BS is inefficient for energy extraction. Furthermore, BSs can attain over-extremal rotation, $1 < \beta \leq 9/8$, beyond the BH extremal limit. Similarly to BHs, the Buchdahl stars can also possess ergospheres for $1/\sqrt{2}< \beta \leq 9/8$, thereby enabling the investigation of their energetics and providing a deeper understanding of their physical properties.  

Therefore, we investigated the magnetic-reconnection process from a Buchdahl star using the Comisso–Asenjo mechanism \cite{Comisso21}. In this framework, energy extraction occurs in the ergoregion via accelerated plasma  with $\epsilon_{+}^{\infty}>0$ and decelerated plasma with the energy $\epsilon_{-}^{\infty}<0$. We found that the energy at infinity per unit enthalpy of the plasma, as well as the energy extraction efficiency, strongly depend on the spin parameter, magnetization parameter, and the orientation angle of the reconnecting magnetic field lines. We further analyzed the dependence of this energy on the magnetization parameter for both accelerated and decelerated plasma components, considering $\beta$=1 (Kerr BH) and $\beta$=9/8 (BS). While the energy of the decelerated plasma exhibits similar behavior for both cases, the accelerated component shows a stronger dependence on magnetization, with a more pronounced increase for the Kerr BH. This difference is also reflected in the energy extraction process. 
 
Furthermore, by examining phase-space regions $\{\beta, r/M\}$ for BS, we showed that magnetic-reconnection-driven energy extraction strongly depends on the orientation angle of the plasma flows in the equatorial plane. In particular, lower values of the magnetization parameter and orientation angle enlarge the phase-space region where energy extraction is possible. 
We further analyzed the power extracted from a maximally rotating BS as a function of the dominant X-point location for different magnetization parameters and orientation angles. Our results show that the extracted power of BS increases with higher magnetization and with a smaller orientation angle between the reconnecting magnetic field lines.

More importantly, we compared the MR energy extraction efficiency for maximally rotating Kerr BH ($\beta$=1) and BS ($\beta$=9/8). We found that MR is more efficient for BS, compared to Kerr BH (see Fig.~\ref{fig:efficiencyeta}). We also analyzed the efficiency distribution across the ergosphere, revealing clear differences between the two cases, as shown in Fig.~\ref{fig:efficiencyplot}. Finally, we evaluated the energy extraction rate in the fast MR regime by comparing the extracted power with that of the BZ mechanism \cite{Blandford1977}. We analyzed the ratio $P_{extr}/P_{BZ}$ as a function of the plasma magnetization for various configurations. Our results showed that for a wide range of magnetization values, MR yields significantly higher power than the BZ process, i.e., $P_{extr}/P_{BZ}>>1$, indicating its higher efficiency.

Our findings show that MR can significantly enhance the energy extraction from rapidly rotating BSs, potentially making them more efficient than BHs as engines of high-energy astrophysical phenomena. Moreover, our analysis offers qualitative insight into the nature of such compact objects and their fundamental distinctions from BHs.

\appendix

\bibliographystyle{apsrev4-1}  
\bibliography{gravreferences,Ref_BS1,Ref_BS2}

\end{document}